\documentclass{mpe_report}

\usepackage{psfig,graphicx,epsfig}
\usepackage{color}
\usepackage{amsmath,amssymb,epic,eepic,array}

\begin{document}

\pagenumbering{arabic}
\setcounter{page}{134}

\renewcommand{\FirstPageOfPaper }{134}\renewcommand{\LastPageOfPaper }{136}

\title{On a synthetical method to constrain 3-D structure of emission regions of pulsars}
\author{H.G. Wang\inst{1}, H. Zhang\inst{2}, X.H. Cui\inst{3}, G.J. Qiao\inst{3}, K.J. Lee\inst{3}, Y. Liu\inst{1} and R.X. Xu\inst{3}}
\institute{Center for Astrophysics, Guangzhou University,
Guangzhou 510006, China \and National Astronomical Observatories,
Chinese Academy of Sciences, Beijing 100012, China\and Department
of Astronomy, Peking University, Beijing 100871, China }
\authorrunning{Wang et al.}
\maketitle

\begin{abstract}
The main idea and procedure are presented for a new synthetical
method on constraining 3-D structure of emission regions of
pulsars. With this method the emission regions can be
synthetically constrained by fitting multi-wavelength observations
features, e.g. pulse widths, phase offsets between radio pulse
profiles and high energy light curves, and radio polarization
properties. The main technique is based on numerically calculating
the emission directions along any open field line via including
aberration and retardation effects under a certain magnetic field
configuration, e.g. vacuum dipole field. It can be easily extended
by involving some further effects or magnetic field line
configuration. The role of observed linear polarization properties
in the method is discussed. The future application to known
X/gamma-ray and radio pulsars will be helpful to test and improve
current theoretical models on pulsar emission.

\end{abstract}

\section{Introduction}

Although pulsars have been studied for 40 years, the structures of
radio and gamma-ray emission regions of pulsars remain
controversial. It is known that radio emission heights are usually
less than 10\% light cylinder  radius for normal pulsars (e.g.
Cordes 1978, Ranin 1990, Blaskiewicz et al. 1991, Kijak \& Gil
1997), but the trans-field structure of radio emission regions,
both in polar or azimuthal directions, were not well identified
from observations in literature.

Finding out the structure of emission regions is helpful to the
study on some theoretical issues, e.g. the structure of radio
emission beam and origin of gamma-ray emission. To explain various
kinds of radio average pulse profiles, some authors proposed that
radio beam is composed of conal and core sub-beams (e.g. Rankin
1983, Gil \& Krawczyk 1996), while some others suggested that
radio beam is just composed of randomly distributed sources
(Manchester 1995, Han \& Manchester 2001). Since emission beam is
formed by radiation of particles from flux tubes in emission
region, the structure of beam is related to the structure of
emission region. Therefore, it is not clear yet whether the radio
emission region is composed of conal and core-like flux tubes or
randomly distributed flux tubes. The controversy for the origin of
gamma-rays remains in different models, i.e. the slot gap model
(e.g. Muslimov \& Harding 2004), the outer gap model (Cheng et al.
2000, Hirotani \& Shibata 1999) and the annular gap model (Qiao et
al. 2004). These models suggest different kind of particle
accelerators and different gamma-ray emission regions. Therefore,
determining the 3-D structure of emission regions from observation
is necessary to test and improve pulsar emission models.

We are developing a synthetical method to constrain the emission
regions from observational properties. The main technique is
numerical calculation for the emission directions along any open
field line by including aberration and retardation effects under a
certain magnetic field configuration, e.g. vacuum dipole field. In
this paper we focus on its main idea and some remaining problems.

\section{The new synthetical Method}

\subsection{The main idea and procedure}

The method uses three basic assumptions. (1) In the co-rotation
reference frame, the emission direction is assumed to be along the
tangent of magnetic field line. The $1/\gamma$ solid angle of
emission cone of a relativistic particle is neglected for the
purpose of simplifying calculation. (2) The emission patterns are
the same at two opposite poles. (3) Vacuum rotating dipolar field
is adopted. In the Cartesian coordinate system of which the
rotating axis is $z$-axis and $x$-axis locates on the meridian
plane, the Cartesian components of the magnetic field vector at
any point P($x, y, z$) are as follows (Cheng et al. 2000)

\begin{eqnarray}
\lefteqn{B_{\rm
x}={\mu\over{r^5}}\left\{3xz\cos\alpha+\sin\alpha\left(
\left[(3x^2-r^2)+3xyr_{\rm n}
\right.\right.\right.}\nonumber\\
&&\left.+(r^2-x^2)r_{\rm n}^2\right]\cos [(R-r)/r_{\rm c}]+\left[3xy-(3x^2-r^2)r_{\rm n}\right. \nonumber \\
&&\left.\left.\left. -xyr_{\rm n}^2\right]\sin [(R-r)/r_{\rm c}]\right)\right\},\nonumber\\
\lefteqn{B_{\rm
y}={\mu\over{r^5}}\left\{3yz\cos\alpha+\sin\alpha\left(
\left[3xy+(3y^2-r^2)r_{\rm n}\right.\right.\right.}\nonumber\\
&&\left.-xyr_{\rm n}^2\right]\cos [(R-r)/r_{\rm c}]+\left[(3y^2-r^2)-3xyr_{\rm n}\right.\nonumber\\
&&\left.\left.\left. +(r^2-y^2)r_{\rm n}^2\right]\sin [(R-r)/r_{\rm c}]\right)\right\},\nonumber\\
\lefteqn{B_{\rm
z}={\mu\over{r^5}}\left\{(3z^2-r^2)\cos\alpha+\sin\alpha\left(
\left[3xz+3yzr_{\rm n}-xzr_{\rm n}^2\right]\right.\right.}\nonumber\\
&&\left.\left.\times\cos [(R-r)/r_{\rm c}]+\left[3yz-3xzr_{\rm
n}-yzr_{\rm n}^2\right]\right.\right.\nonumber\\
&&\left.\left.\times\sin [(R-r)/r_{\rm c}]\right)\right\}.
\label{bret}
\end{eqnarray}
where $\mu$ is the magnetic moment of the pulsar, $R$ is stellar
radius, $r$ is the emission radius from stellar center to P,
$r_{\bf n} = r/r_{\rm c}$, $r_{\rm c}$ is the radius of light
cylinder, $\alpha$ is inclination angle between rotation and
magnetic axes.

The method runs in two steps as shown schematically in Fig. 1.
{\it Step 1, determine geometrical parameters of the pulsar and
find out all emission points that can be viewed by our line of
sight (hereafter LOS).} To perform the calculation, the whole open
volume is divided into a number of layers from the last open field
lines to the magnetic axis (see Fig. 1 for the shape of foot
points of the field lines in each layer on polar cap). (1)
Determine or constrain the values of $\alpha$ and $\zeta$, where
$\zeta$ is the viewing angle between line of sight and rotation
axis. Definite values of $\alpha$ and $\zeta$ will greatly reduce
calculation. For simplification one can use the values that are
obtained by fitting the observed position angle (PA) of linear
polarization with rotation vector model (hereafter RVM,
Radhakrishnan \& Cooke 1969). But strictly to say they should be
figured out consistently with the method, because aberration,
retardation and non-static dipole field configuration all
influence position angle. (2) Calculate tangents of field lines
and then modify emission direction by involving aberration effect.
To do the modification the initial emission direction (tangential
to field line) in the co-rotation reference frame is converted to
the direction in laboratory reference frame via Lorentz transform.
(3) Select all the emission points where the emission direction is
aligned with our LOS.

{\it Step 2, find out the emission points that contribute to
emission by fitting the observed multi-wavelength properties.} The
way to do this is based on $\Phi-\phi$ diagram, in which the
$\Phi-\phi$ curves for different layers of field lines are
plotted, where $\Phi$ is the pulse longitude of the emission from
a point on a field line and $\phi$ is azimuthal angle of the foot
point of the field line. After figuring out the diagram, one can
place the observed multi-wavelength pulse windows on the diagram,
and then finds out which field lines may contribute to the pulse
(see lower panel in Fig.1). In some cases linear polarization
percentage may provides further constraints on the emission
region. The role of polarization properties in this method is
discussed in next section.

For a typical pulsar that has radio and high energy emission the
observed properties used include (a) the radio and high-energy
pulse widths, (b) the phase separation between radio main pulse
and inter-pulse (in the case that inter-pulse is observed), (c)
the phase offset between radio and high energy pulses, (d) radio
polarization properties, viz. linear polarization percentage and
PA sweep, (e) viewing angle determined from the elliptical shape
of torus or rings in nebula for some pulsars, e.g the Crab and
Vela pulsars.
\begin{figure}
\centerline{\psfig{file=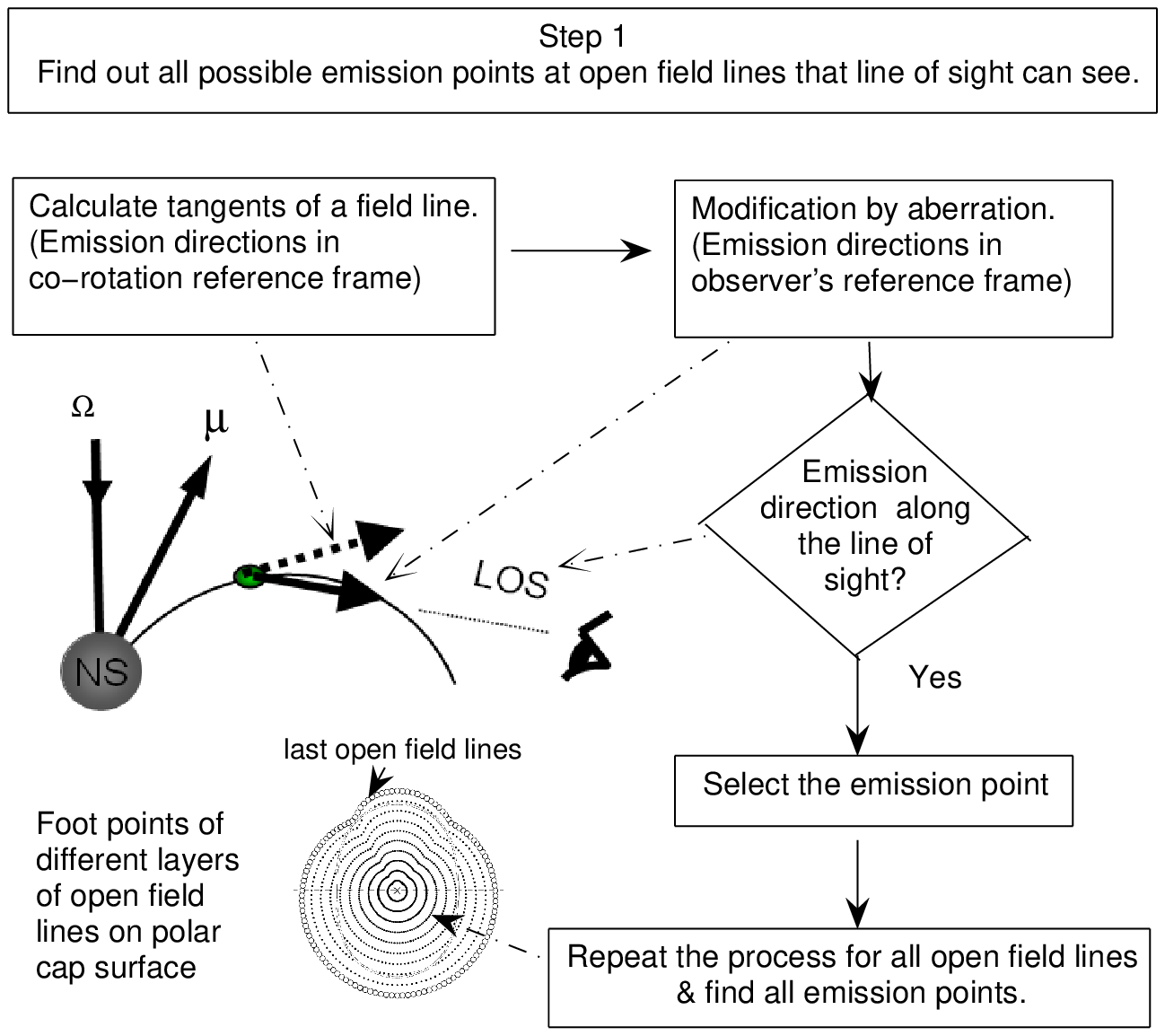,width=8.8cm,clip=} }
\centerline{\psfig{file=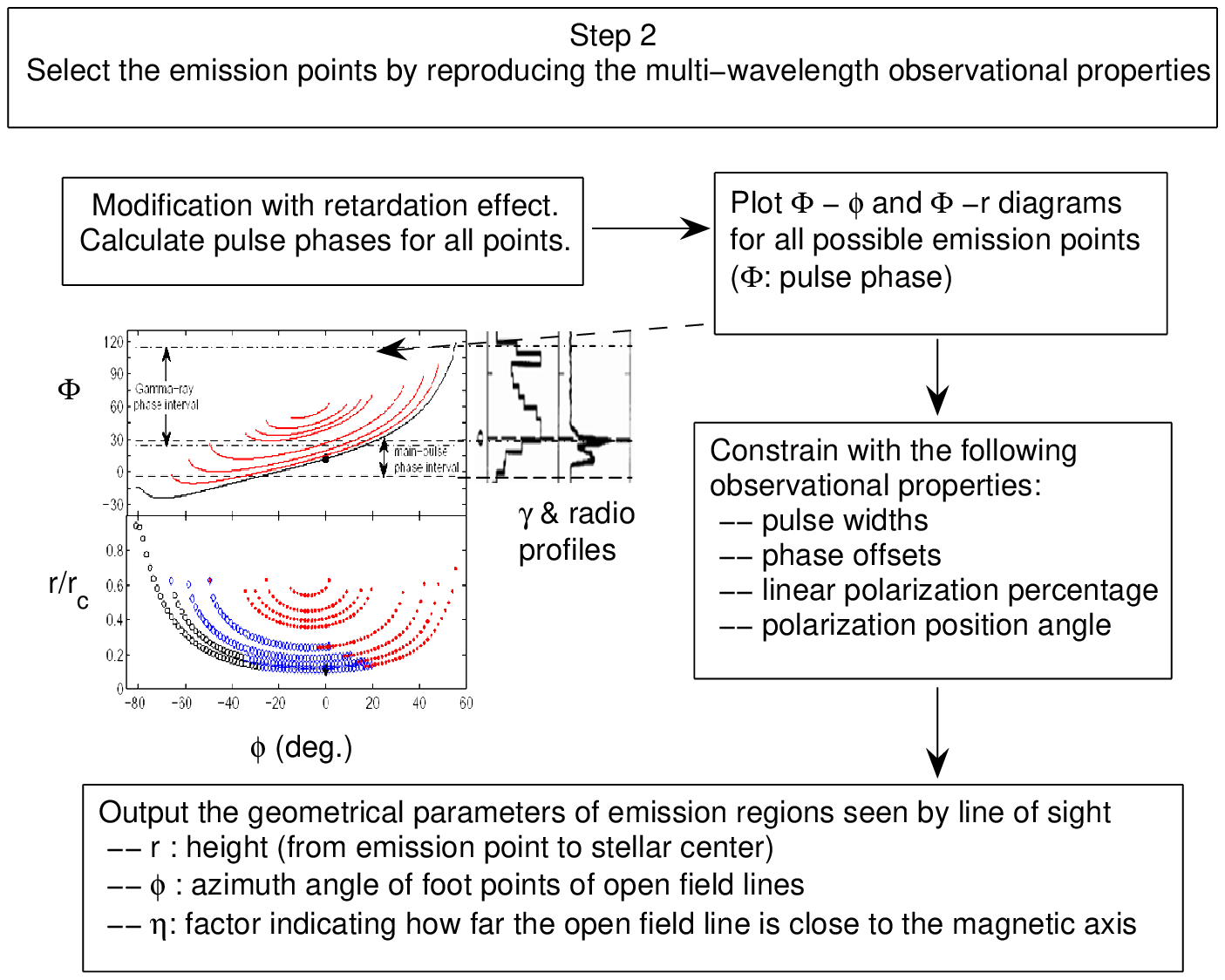,width=8.8cm,clip=} }
\caption{Illustration to the main procedure of the synthetical
method. \label{fig1}}
\end{figure}
\subsection{Comparison with conventional methods}

Comparing with conventional geometrical and relativistic models to
constrain emission altitudes, our new method has three advantages.
(1) It can apply to high emission altitudes and complex magnetic
field configuration. (2) It can reveal asymmetrical structure of
emission regions. (3) It is more synthetical. Details are as
follows.

Geometrical methods (Rankin 1990, Kijak \& Gil 1997, Wu et al.
2002) and relativistic methods (Blaskiewicz et al. 1991, Dyks et
al. 2004, Gangadhara 2005) are two main conventional approaches to
constrain the radio emission heights. In geometrical methods
geometry relations of emission beam and pure dipole field are
used, while in relativistic model the phase-delay-to-radius
relation is used by considering the aberration and retardation
effects in static dipolar field. But geometrical relations and
approximations of aberration and retardation effects are only
valid when emission heights are not too large. However, in
numerical method such limitation vanishes, because reduction of
aberration and retardation effects in complex field line
configuration can be easily done by vector operations. Therefore,
numerical method can be used to constrain the gamma-ray emission
region and radio emission region of some young pulsars where the
emission height may be close to light cylinder.

A common feature of conventional geometrical and relativistic
methods is that the emission region is assumed to be symmetrical
with respect to the meridian that contains the rotation and
magnetic axes. However, the effects of aberration, retardation and
sweeping back of magnetic fields probably make the emission region
seen by LOS asymmetrical to the meridian. In our method the
assumptions are relaxed, so that asymmetrical properties of
emission regions may be revealed.

From intuition, more information may be revealed by constraining
emission regions synthetically with multi-wavelength emission
properties than with only radio or gamma-ray properties. It has
been shown that some interesting results can be obtained for radio
and gamma-ray emission regions of PSR B1055-52 by synthetically
reproducing the radio and gamma-ray properties, even with a
modified geometrical method (Wang et al. 2006). The new
synthetical method is hopeful to go further.

\section{Discussion}

Besides the pulse width and phase offsets, linear polarization is
another key factor in this synthetical method. What results one
finally gets depend on how to use the polarization data. The
polarization can be used in two ways. First, the PA data should be
self-consistently fitted with the method to solve or constrain
$\alpha$, $\zeta$ and the range of layers of open field lines. For
some pulsars with nebulae, e.g. the Crab and Vela pulsars, the
viewing angle can be determined from the torus and ring-like
structure near pulsar. In this case the uncertainty in geometry is
much reduced. Second, linear polarization percentage (hereafter
LPP) can be used to constrain the polar and azimuthal scales of
emission region. (1) In the cases of nearly 100\% LPP, the
emission region should be limited within a thin layer of flux
tube, otherwise the superposition of radio waves from different
layers, which have different PAs, may cause significant
depolarization. (2) In the case that LPP is much less than 100\%,
it can not be simply concluded that the emission region is thick
in polar scale, because depolarization may be caused by either
superposition of emissions with scattered PAs or superposition of
orthogonal modes (Stinebring et al. 1984). Detailed studies on PA
distribution of single pulses may be helpful to distinguish them.

A realistic magnetic field is different from the vacuum rotating
dipole field. Both magnetospheric current and general relativistic
effect can influence the field configuration. It would be
interesting to test how the constrained results depends on field
configuration.

Our future work includes constraining (1) gamma-ray, X-ray and
radio emission regions of gamma-ray pulsars, (2) X-ray and radio
emission regions for X-ray pulsars, (3) radio emission regions for
radio pulsars with high linear polarized percentage and
good-quality data of position angle.

\begin{acknowledgements}
We are grateful to R.N. Manchester for his valuable discussion.
This work is supported by NSF of China (10373002, 10403001,
10573002)
\end{acknowledgements}

       \clearpage

\end{document}